\documentclass[%
prx,
 amsmath,amssymb,
 aps,
twocolumn,
longbibliography,
showkeys
]{revtex4-1}

\usepackage{graphicx}
\usepackage{epstopdf}
\usepackage{dcolumn}
\usepackage{bm}
\usepackage{color}
\usepackage[mathlines]{lineno}
\usepackage[breaklinks=true,colorlinks=true,linkcolor=blue,urlcolor=blue,citecolor=blue]{hyperref}
\usepackage{ragged2e}
\usepackage{enumitem}
\usepackage{bibunits}
\usepackage{amsthm}
\usepackage{diagbox}
\usepackage{booktabs}

\renewcommand\thefigure{\arabic{figure}}

\begin{document}


\title[]{Indirect punishment can outperform direct punishment in promoting cooperation in structured populations
}

\author{Yujia Wen$^1$}
\author{Zhixue He$^{2,1}$}
\author{Chen Shen$^3$}
\email{steven\_shen91@hotmail.com}
\author{Jun Tanimoto$^{1,3}$}
\email{tanimoto@cm.kyushu-u.ac.jp}

\affiliation{
\vspace{2mm}
\mbox{1. Interdisciplinary Graduate School of Engineering Sciences, Kyushu University, Fukuoka, 816-8580, Japan}
\mbox{2. School of Statistics and Mathematics, Yunnan University of Finance and Economics, Kunming, 650221, China}
\mbox{3. Faculty of Engineering Sciences, Kyushu University, Fukuoka 816-8580, Japan}
}

\date{\today}

\begin{abstract}
Indirect punishment traditionally sustains cooperation in social systems through reputation or norms, often by reducing defectors’ payoffs indirectly. In this study, we redefine indirect punishment for structured populations as a spatially explicit mechanism, where individuals on a square lattice target second-order defectors---those harming their neighbors---rather than their own immediate defectors, guided by the principle: ``I help you by punishing those who defect against you''. Using evolutionary simulations, we compare this adapted indirect punishment to direct punishment, where individuals punish immediate defectors. Results show that within a narrow range of low punishment costs and fines, adapted indirect punishment outperforms direct punishment in promoting cooperation. However, outside this cost-fine region, outcomes vary: direct punishment may excel, both may be equally effective, or neither improves cooperation, depending on the parameter values. These findings hold even when network reciprocity alone does not support cooperation. Notably, when adapted indirect punishment outperforms direct punishment in promoting cooperation, defectors face stricter penalties without appreciably increasing punishers’ costs, making it more efficient than direct punishment. Overall, our findings provide insights into the role of indirect punishment in structured populations and highlight its importance in understanding the evolution of cooperation.
\end{abstract}

\keywords{cooperation, punishment, spatial prisoner's dilemma game}

\maketitle

\section*{\label{sec:level1}Introduction}

The advancement of society relies on cooperation among individuals to maximize collective benefits~\cite{axelrod1981evolution,ohtsuki2006simple,nowak2006five,boyd2009culture}. However, cooperation imposes individual costs while benefiting others, making it vulnerable to exploitation by selfish free-riders~\cite{hardin2018tragedy}. Understanding how cooperation emerges among strangers, particularly in one-shot and anonymous encounters~\cite{fehr2002altruistic}, is crucial because these settings artificially remove key mechanisms that typically sustain cooperation. Such mechanisms include repeated interactions~\cite{camerer2011behavioral,hilbe2018partners,hu2019modelling,zhu2022nash}, reputation effects~\cite{xia2023reputation,nowak1998evolution}, and other reciprocity mechanisms~\cite{rand2013human}, creating the most challenging conditions for cooperation to emerge. Yet, despite these obstacles, economic game experiments consistently show that people do cooperate, with cooperation levels stabilizing at lower but persistent rates~\cite{chaudhuri2011sustaining,andreoni1999preplay}. Evidence suggests that cooperation in such scenarios may result from prosocial preferences or confusion about the game structure\cite{fehr2002strong,wang2024confusion,burton2021payoff,shen2024beyond}. Nevertheless, researchers continue to explore mechanisms to sustain cooperation, particularly through costly punishment\cite{dreber2008winners,fehr2002altruistic}. Beyond human interactions, punishment is widely observed in nature~\cite{clutton1995punishment,raihani2012punishment}. In animal populations, individuals often use punishment to constrain non-cooperative or deceptive behaviors, thereby promoting population stability~\cite{riehl2016cheating}. These findings suggest that punishment is a fundamental enforcement mechanism shaping cooperative behavior across biological and social systems.

Costly punishment generally takes two forms: direct punishment and indirect punishment. Direct punishment occurs when an individual penalizes a defector who has personally harmed them~\cite{dreber2008winners,wu2009costly,li2018punishment,zhu2023equilibrium}. While this strategy can deter defection, it imposes a cost on the punishers and risks retaliation, making it evolutionarily unstable unless supported by additional mechanisms such as reputation effects~\cite{santos2011evolution}, voluntary participation~\cite{fowler2005altruistic}, or network reciprocity~\cite{perc2012self,szolnoki2017second,guo2023self,helbing2010evolutionary,fu2017leveraging,wang2024deterministic,chen2024new}. By contrast, indirect punishment occurs when an individual punishes a defector who has harmed someone else, even though the punisher was not personally affected. This enforcement mechanism helps maintain cooperation by discouraging selfish behavior on a broader social scale. A well-documented example is indirect reciprocity, where individuals refuse to cooperate with those who have previously defected against someone else~\cite{nowak1998evolution,rockenbach2006efficient,ule2009indirect,balafoutas2014direct}. Experimental studies suggest that even when indirect punishment is rarely used, its mere availability can sustain cooperative behavior by reinforcing prosocial norms~\cite{ule2009indirect}. Unlike direct punishment, which requires personal retaliation, indirect punishment typically relies on social mechanisms such as reputation systems to track defectors’ past actions and coordinate sanctions~\cite{nowak1998evolution,xia2023reputation,balafoutas2014direct}. While it may lower retaliation risks, it still incurs costs, such as monitoring effort and the risk of misjudging defectors.

Owing to these differences, the evolutionary dynamics of direct punishment have been extensively studied in structured populations, where network structures stabilize it by balancing second-order free-riding (cooperators who don’t punish) and antisocial punishment (targeting cooperators), thus ensuring their survival~\cite{perc2012self,szolnoki2017second,helbing2010evolutionary}. By contrast, indirect punishment has received less attention in spatial networks, as it traditionally relies on reputation or norms to sanction defectors across broader social scales, often among strangers~\cite{ule2009indirect,balafoutas2014direct}. In structured populations where interactions are localized, direct punishment can naturally be applied to nearby defectors. However, monitoring reputations and enforcing norms can be challenging, potentially weakening the effectiveness of aforementioned indirect punishment within such populations. 
Despite these constraints, the fundamental principle of indirect punishment---sanctioning defectors for others’ sake---suggests that alternative spatially adapted mechanisms, independent of global reputation, may still facilitate cooperation. This contrast raises a pivotal question: in structured populations, how does spatially adapted indirect punishment compare to direct punishment in sustaining cooperation, and under what conditions might it prove more effective?

To address this gap, we examined the effectiveness of direct and indirect punishment in structured populations using an extended evolutionary prisoner’s dilemma game on a square lattice.  The game consists of two stages: in the first stage, players decide whether to cooperate with their direct neighbors, and in the second stage, they choose whether to impose fine on defectors at a personal cost. Direct punishment is implemented by allowing players to punish their immediate neighbors (depicted as the purple-shaded area in Fig.\ref{Fig_intro}). Meanwhile, indirect punishment is adapted to the square lattice, requiring focal players to punish their indirect neighbors, specifically the neighbors of their neighbors (illustrated as the cyan-shaded area in Fig.~\ref{Fig_intro}). Although this adaptation differs from traditional reputation-based indirect punishment models, it maintains the core idea of indirect punishment, where defectors are sanctioned even if they have not directly harmed the punisher.

Through evolutionary simulations, our results show that under conditions of low punishment costs and fines, indirect punishment promotes cooperation more effectively than direct punishment. Outside this region, when punishment costs are low but fines are high, both punishment forms can be equally effective. When the cost and fine are relatively  high, direct punishment becomes more favorable in other cases. However, when the cost is high and fine is low, neither strategy significantly enhances cooperation. These results are robust in terms of when network reciprocity can and cannot promote cooperation. In scenarios where indirect punishment outperforms direct punishment, defectors encounter stricter penalties without substantially increasing the costs borne by punishers, making it a more efficient enforcement mechanism. These findings suggest that under specific low-cost conditions, indirect punishment serves as a more effective method for promoting cooperation than direct punishment, offering valuable insights into optimizing punishment mechanisms within structured populations.

\section*{Model and method}

Considering a population of size $N$ placed on a square lattice with moore neighborhood and periodic boundaries, in which players play two-stage prisoner's dilemma game (PDG) with their neighbors \cite{szabo2007evolutionary}. In the first stage of the PDG, paired players simultaneously choose to either cooperate or defect, with the outcomes determined by the actions of both parties. Specifically, players can obtain payoff $R$ from mutual cooperation and a $P$ from mutual defection. When unilateral defection occurs, cooperator receives $S$ while defector receives $T$. The social dilemma arises when the payoff elements satisfy $T>R>P>S$ and $2R>T+S$, indicating that while mutual cooperation maximizes collective benefits, defection provides the highest individual benefit. To simplify the model without losing generality, we set $R=1$ and $P = 0$, and follow ref.\cite{wang2015universal} to introduce the dilemma strength parameter $r:= \frac{T-R}{R-P} = \frac{P-S}{R-P} $, which characterizes the relative benefit of defection over cooperation. Thus, $T=1+r$ and $S=-r$.

\begin{table*}[!t]
  \caption{ \label{tab:payoff_matrix} 
 The payoff matrix for the two-stage Prisoner's Dilemma game: (a) The first-stage interaction involves players deciding whether to cooperate; (b) The second stage involves players deciding whether to punish, with the interaction outcomes under different strategies representing the row player's payoff in that stage.
 }
    \centering
\begin{minipage}[t]{0.30\linewidth}
\resizebox{0.99\linewidth}{!}{
\begin{tabular}{p{0.1\linewidth} p{0.2\linewidth} p{0.2\linewidth} p{0.2\linewidth} p{0.2\linewidth}}
\multicolumn{5}{c}{(a) The first-stage interaction}  \\
\hline
 & $CN$     & $DN$     & $CP$    & $DP$    \\
\hline
$CN$  & 1 & -$r$  & 1  & -$r$ \\
$DN$  & 1+$r$  & 0  &  1+$r$  & 0 \\
$CP$  & 1 & -$r$  & 1  & -$r$ \\
$DP$  & 1+$r$  & 0  &  1+$r$  & 0 \\
\hline
\end{tabular}
}
\end{minipage}
\begin{minipage}[t]{0.30\linewidth}
\resizebox{0.99\linewidth}{!}{
\begin{tabular}{p{0.1\linewidth} p{0.2\linewidth} p{0.2\linewidth} p{0.2\linewidth} p{0.2\linewidth}}
\multicolumn{5}{c}{(b) The second-stage interaction}  \\
\hline
 & $CN$     & $DN$     & $CP$    & $DP$    \\
\hline
$CN$  & 0 & 0  & 0  & 0 \\
$DN$  & 0  & 0  & -$\beta$  & -$\beta$ \\
$CP$  & 0  & -$\gamma$  & 0  & -$\gamma$ \\
$DP$  & 0  & -$\gamma$  & -$\beta$  & -$\beta$ -$\gamma$ \\
\hline
\end{tabular}
}
\end{minipage}
\label{ts1}
\end{table*}


In the second stage, players decide whether to punish defectors based on first-stage outcomes. Punishment involves punishers incurring a cost ($\gamma$) to impose a fine ($\beta$) to others. Based on first-stage actions, we define four strategies: cooperating without punishing ($CN$), defecting without punishing ($DN$), cooperating and punishing defectors ($CP$), and defecting while punishing defectors ($DP$). The payoffs for these four strategies are summarized in Table~\ref{tab:payoff_matrix}.

\begin{figure}[!t]
    \centering
    \includegraphics[width=0.80\linewidth]{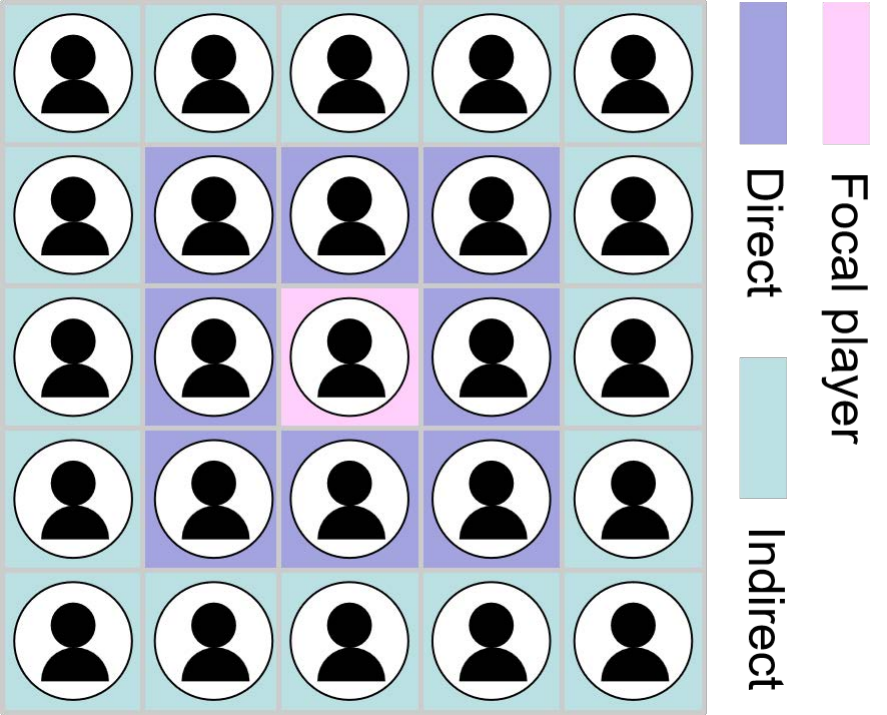}
      \caption{\textbf{Schematic diagram of direct and indirect punishment in the model.}  In the first stage of the prisoner’s dilemma game, a focal player interacts with their eight direct neighbors (purple-shaded area). In the second stage, punishment targets differ: direct punishment imposes penalties on defective direct neighbors, while indirect punishment targets defective second-order neighbors (up to sixteen, cyan-shaded area), excluding direct neighbors.
     }
    \label{Fig_intro}
\end{figure}

In direct punishment, a focal player interacts with their direct neighbors in both stages (highlighted in purple in Fig.~\ref{Fig_intro}). In contrast, indirect punishment differs in that second-stage interactions target second-order neighbors (i.e., neighbors of first-order neighbors, highlighted in cyan in Fig.~\ref{Fig_intro}). Notably, punishing second-order defectors ultimately benefits first-order neighbors, embodying the principle: ``I help you by punishing those who defect against you". From this perspective, this form of punishment can be viewed as an indirect reward to direct neighbors.

The total payoff of a player is the sum of the outcomes from the two-stage interactions. Specifically, in the case of direct punishment, the focal player's payoffs in the first and second stages are calculated according to Table \ref{tab:payoff_matrix} (a) and (b), respectively, based on interactions with eight direct neighbors. In the case of indirect punishment, the focal player's payoff in the first stage is calculated through interactions with eight immediate neighbors according to Table \ref{tab:payoff_matrix} (a), while the second-stage payoff is obtained through interactions with sixteen second-order neighbors according to Table \ref{tab:payoff_matrix} (b).


We use Monte Carlo simulations to obtain the results for both the direct punishment and indirect reward models. Initially, players randomly adopt one of the four strategies. In each time step of the Monte Carlo simulation, $N$ updates occur. During each update, a focal player is randomly selected, and their payoff is calculated according to the above rules, the player then updates strategy through pairwise imitation. Thus, on average, each player updates strategy once per time step. The focal player $i$ mimics the strategy of a randomly selected direct neighbor $j$ with a Fermi probability\cite{sigmund2010social}:
\begin{equation}
    w_{i \leftarrow j}=\frac{1}{1 + \exp \{ {(\Pi_i-\Pi_j)\kappa } \} },
\end{equation}
where $\Pi_{x}$ represents the player $x$'s total payoffs. The parameter $\kappa$ denotes the imitation strength; as $\kappa \to 0$, imitation approaches randomness, while as $\kappa \to+\infty$ , imitation is driven by payoff differences. In this study, we set $\kappa=10$ to investigate a strong imitation scenario. The network sizes $N$ range from $100^2$ to $400^2$ in simulation, and the final results are averaged from over 50 independent simulations, with each simulation producing average values from the last 2000 time steps out of a total exceeding $1\times10^5$ time steps.

\section*{Results}

In structured populations, network reciprocity can enable cooperative clusters to resist defectors~\cite{nowak2006five, wang2013insight}, thereby promoting cooperation. However, under high dilemma strength, network reciprocity fails to prevent defectors from infiltrating cooperative clusters, making cooperation difficult to establish. To investigate the role of punishment in promoting cooperation, we examine two representative scenarios: low dilemma strength (e.g., $r = 0.05$), where network reciprocity can promote cooperation, and high dilemma strength (e.g., $r = 0.2$), where network reciprocity cannot sustain cooperation.


\begin{figure}[!t]
    \centering
    \includegraphics[width=0.99\linewidth]{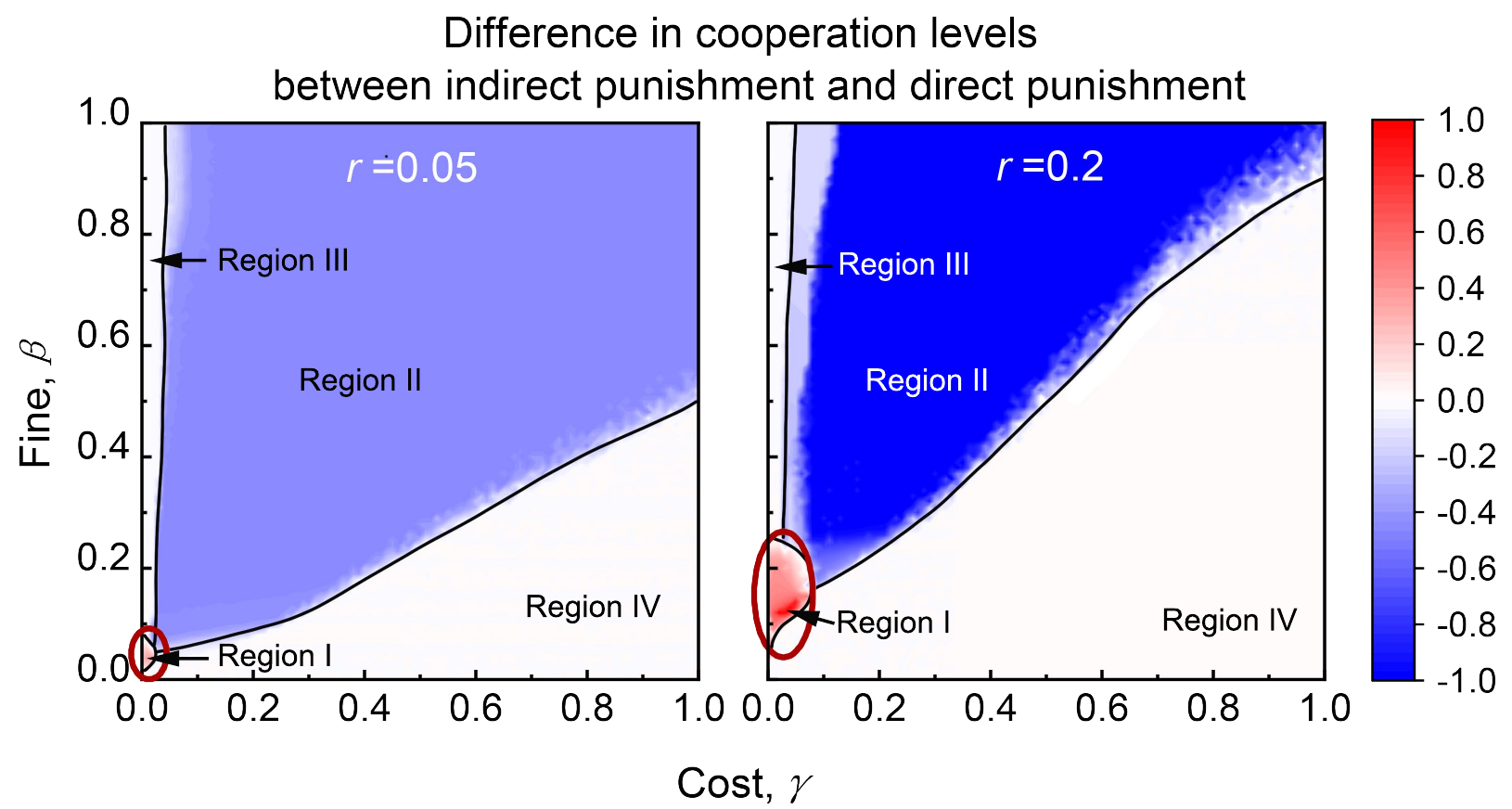}
      \caption{\textbf{Indirect punishment can promote cooperation more effectively than direct punishment under conditions of low-costs and low-fines (marked with dark red circles).} Color codes illustrate the difference in cooperation levels (i.e., the fraction of $CN+CP$ within the population) between direct punishment and indirect punishment as functions of punishment cost $\gamma$ and fine $\beta$ for low dilemma strength (left panel) and high dilemma strength (right panel). Based on the comparison of indirect punishment to direct punishment in promoting cooperation, the parameter space can be divided into the following regions: Region I, where indirect punishment is superior to direct punishment; Region II, where direct punishment is superior to indirect punishment;  Region III and region IV, where the difference between direct and indirect punishment is negligible.}
      \label{Fig_diff}
\end{figure}

\begin{figure}[!t]
    \centering
    \includegraphics[width=0.95\linewidth]{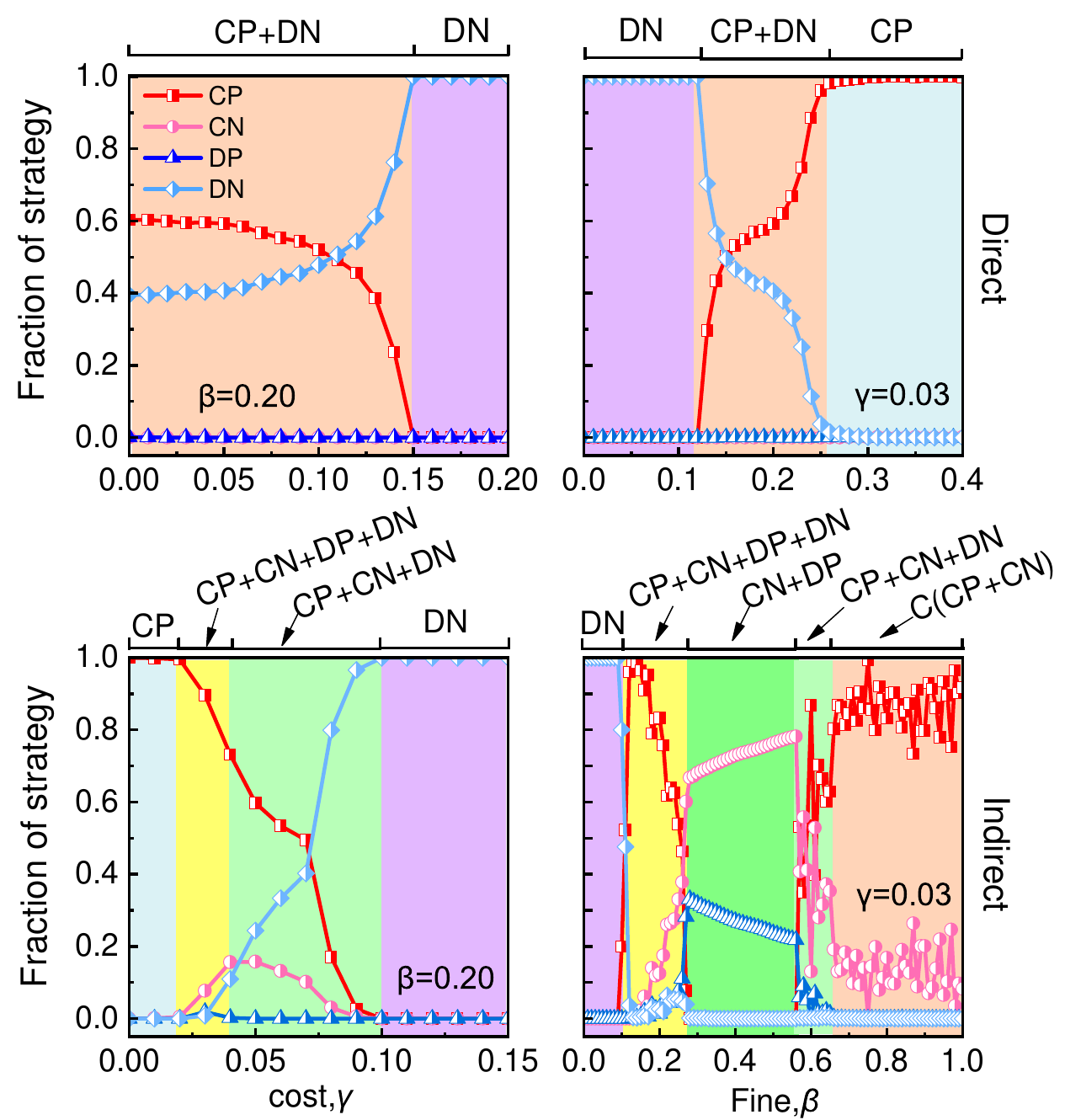}
      \caption{\textbf{
      The effectiveness of indirect punishment in promoting cooperation is sensitive to the costs of punishment.} Depicted are the fraction of $CN$, $DN$, $CP$ and $DP$ as functions of punishment cost $\gamma$ (left panels) and fine $\beta$ (right panels) for direct punishment and indirect punishment. Results are obtained under high dilemma strength. Other parameters are set to $\beta=0.2$ for left panels, and $\gamma$ for right panels.}
    \label{Fig_cross}
\end{figure}

\begin{figure*}[!t]
    \centering
    \includegraphics[width=0.95\linewidth]{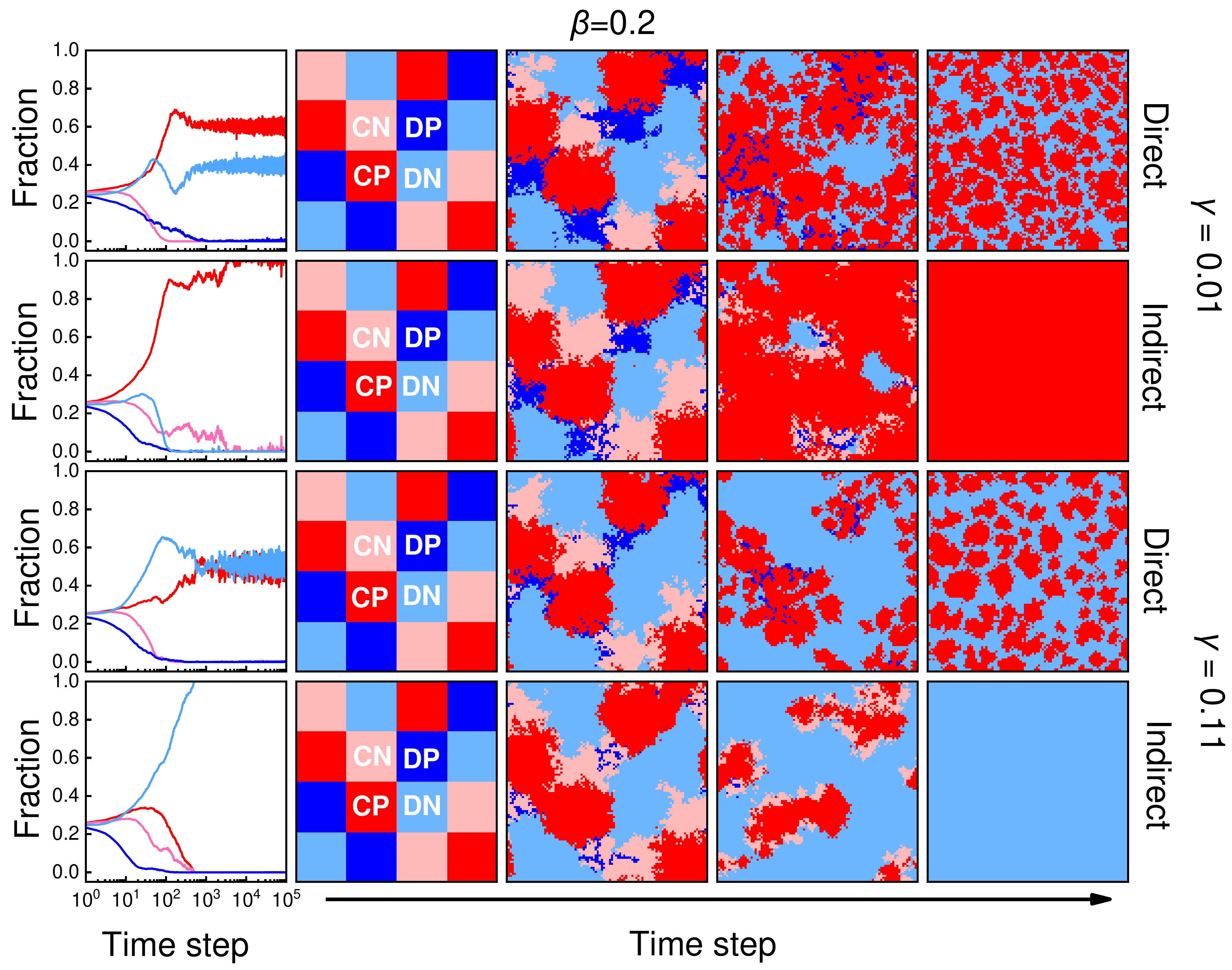}
      \caption{\textbf{Indirect punishment promote the expansion of $CP$ clusters when the punishment cost is low, but lead to their collapse when the cost becomes high}. The leftmost panels illustrate the proportion of the four strategies as a function of time steps, while the right panels present typical evolutionary snapshots starting from a block-like initial distribution for direct punishment and indirect punishment. Light red, dark red, light blue, and dark blue represent $CN$, $CP$, $DN$, and $DP$, respectively. The results are obtained under fixed $\beta = 0.2$ and $r = 0.2$.}
      \label{Fig_snap}
\end{figure*}

Interestingly, indirect punishment can promote cooperation more effectively than direct punishment within a narrow region of punishment parameters. 
The results in Fig.~\ref{Fig_diff} indicate four distinct regions in the parameter space of punishment costs and fines, revealing the differences between indirect and direct punishment in promoting cooperation, regardless of dilemma strength. In region I, characterized by low-costs and low-fines relative to payoff of mutual cooperation, with the fine slightly higher than the punishment cost, indirect punishment can outperform direct punishment, as shown in the red circle of Fig.~\ref{Fig_diff}. In region II, where both costs and fines are relatively high than that of region I, direct punishment is more effective than indirect punishment. In region III, characterized by low-costs and high-fines, both indirect and direct punishment perform similarly, promoting full cooperation, as seen in Appendix Fig.~\ref{figA1} and ~\ref{figA2}. In region IV, characterized by high-costs and low-fines, both punishment methods perform similarly, as neither effectively promotes cooperation.


\begin{figure*}[!t]
    \centering
    \includegraphics[width=0.95\linewidth]{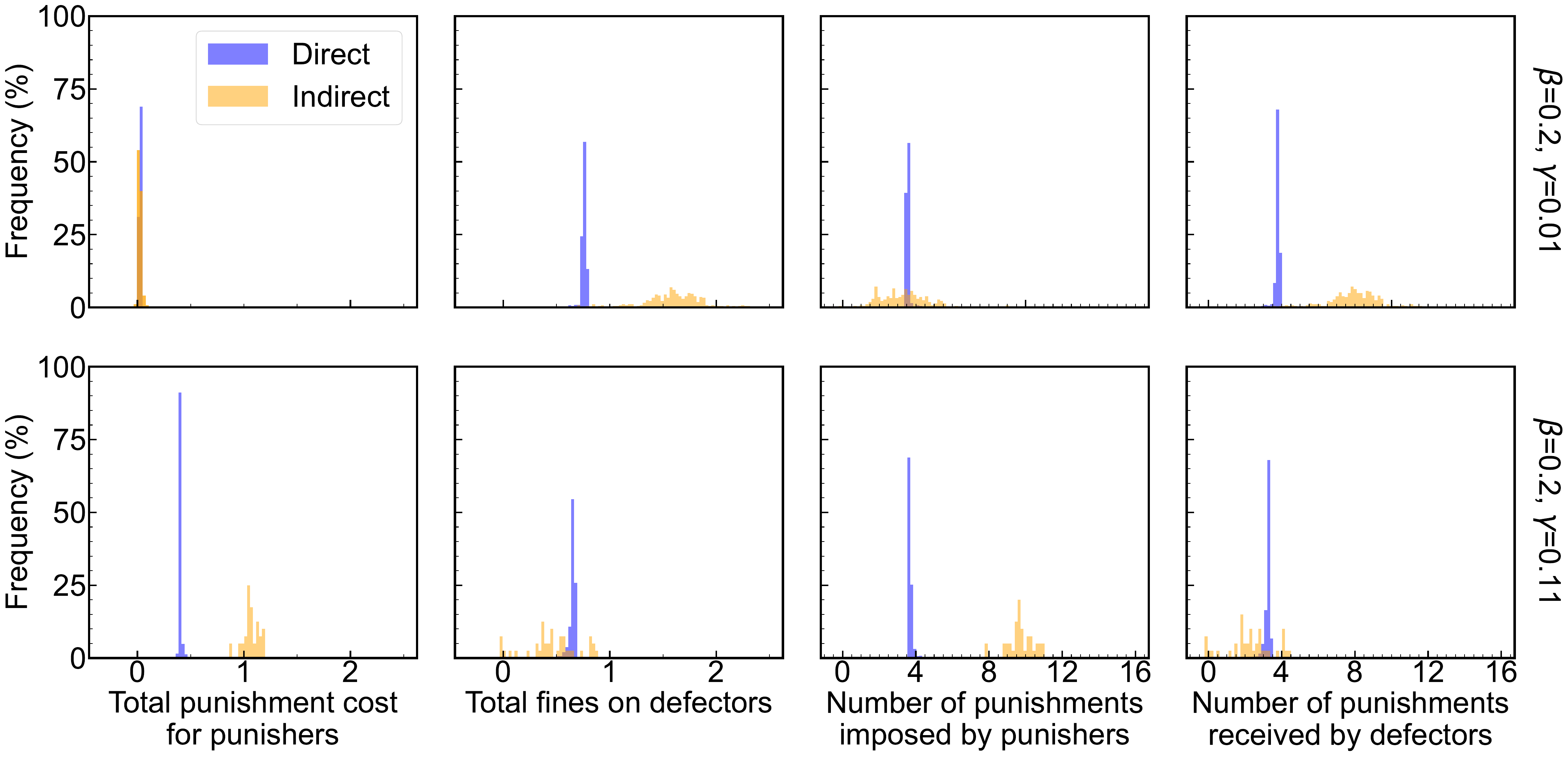}
      \caption{
      \textbf{
      Indirect punishment behavior can lead to defectors being punished more without significantly increasing the total cost for punishers when the punishment cost is low, making them more effective than direct punishment in promoting cooperation. However, high punishment costs result in a greater total punishment cost in the case of indirect punishment, which diminishes the payoff advantage for punishers, making them less effective than direct punishment}. Depicted are the frequency distributions of the total punishment cost for punishers (first column panels from the left), the total fines imposed on the defectors (second column panels), the number of punishments imposed by punishers (third column panels), and the number of punishments received by defectors who are punished (fourth column panels) for direct punishment and indirect punishment. To analyze the evolutionary characteristics of punishers, the distribution data include all evolutionary time steps prior to the stabilization of strategy proportions within the population, and the evolutionary processes correspond to the results presented in Fig.~\ref{Fig_snap}.}
    \label{Fig_dis1}
\end{figure*}

Indirect punishment are more sensitive to changes in punishment costs than direct punishment, which affects the overall efficiency of the punishment mechanism. At a certain fine level (e.g., $\beta = 0.2$), as the cost of punishment increases, it typically reduces the effectiveness of direct punishment. This is evident in the upper left of Fig.~\ref{Fig_cross}, where the proportion of $CP$ gradually declines as $\gamma$ increases, ultimately resulting in the collapse of cooperation. For $\gamma < 0.04$, the proportion of cooperators (i.e., $CN+CP$) under indirect punishment can exceed that under direct punishment, as shown in the lower left of Fig.~\ref{Fig_cross}. Specifically, indirect punishment induces pure cooperation within the population for $\gamma < 0.02$. However, as $\gamma$ continues to rise, the efficiency of indirect punishment declines sharply, leading to the disappearance of cooperation when $\gamma > 0.1$. In contrast, cooperation in the direct punishment scenario only disappears when $\gamma > 0.15 $.

At a given punishment cost level (e.g., $\gamma = 0.03$ ), as the fine increases, direct punishment induces the spontaneous emergence of $CP$ and gradually dominates the population when $\beta > 0.12$, as shown in the upper-right panel of Fig.~\ref{Fig_cross}. In contrast, under indirect punishment, cooperation can be established when $\beta = 0.1$. However, as $\beta$ increases, second-order free riders (i.e., $CN$, who benefit from others' punishments without bearing the cost) gradually replace $CP$. Cooperation reemerges only when $\beta > 0.58$. Additionally, in indirect punishment, the fine required to eliminate defection is $\beta = 0.66$, significantly higher than the $\beta = 0.26$ required for direct punishment. This suggests that indirect punishment requires a larger fine to achieve the same effect as direct punishment.

We then investigate why indirect punishment can be more or less effective than direct punishment in promoting cooperation by analyzing the characteristics of population evolution. To do so, we fix $\beta = 0.2$ and employ a block-like initial distribution to better understand the competitive dynamics between strategies, as shown in Fig.~\ref{Fig_snap}. Under low-cost conditions (e.g., $\gamma = 0.01$), whether through direct punishment or indirect punishment, $DN$ can easily invade $CN$ and $DP$ in the direct punishment scenario, as $DN$ individuals free-ride without bearing the costs of punishment.  It is evident that the proportion of $CP$ steadily increases from the early stages of evolution, indicating that the implementation of punitive measures by $CP$ allows them to resist invasion by $DN$ and contributes to their dominance. Eventually, a coexistence of $CP$ and $DN$ emerges under direct punishment, as shown in the first row of Fig.~\ref{Fig_snap}. In contrast, indirect punishment can support the further expansion of $CP$ clusters, see the second row of Fig.~\ref{Fig_snap}. When $DN$ clusters are surrounded by $CP$, $DN$ on the edge of these clusters, encounters more second-order punishment neighbors, resulting in they suffer more punishment, see the upper panels of Fig.~\ref{Fig_dis1}. Although $CP$ impose more punishment under the indirect punishment, the total punishment cost for $CP$ does not increase significantly due to the lower cost associated with punishment action. By increasing the frequency of punishment on defectors without substantially raising punishers' costs at low $\gamma$, indirect punishment allow $CP$ to outperform $DN$. As a result, indirect punishment can be more efficient than direct punishment in promoting cooperation.

For a high cost (e.g., $\gamma = 0.11$), the size of $CP$ clusters shrinks under direct punishment, but $CP$ can still resist invasion of $DN$, as shown in the third row of Fig.~\ref{Fig_snap}. In contrast, indirect punishment fail to prevent the invasion of $DN$, as shown in the fourth row of Fig.~\ref{Fig_snap}. When $CP$ clusters are surrounded by $DN$, $CP$ individuals have more second-order defecting neighbors, leading to an increase in punitive actions. However, the rise in the total punishment cost limits the expansion of $CP$. Furthermore, despite the increase in punishment imposed by $CP$, these punishments are not effectively concentrated on defectors at the cluster boundaries. As shown in the lower panel of Fig.~\ref{Fig_dis1}, the total fines of $DN$ imposed do not increase significantly. Consequently, indirect punishment fail to sustain cooperation.


\section*{Conclusion and discussion}

To conclude, in this study, we introduced a novel form of indirect punishment on a square lattice, where players impose penalties solely on defective second-order neighbors (the neighbors of their direct neighbors). This spatially adapted strategy indirectly benefits direct neighbors through a spillover effect, enhancing cooperation within the network. Our results demonstrate that indirect punishment outperforms direct punishment---which targets defective direct neighbors---in fostering cooperation under low-cost and low-fine parameter conditions (i.e., region I in Fig.~\ref{Fig_diff}). In low-cost, high-fine conditions (see region III in Fig.~\ref{Fig_diff}), both strategies are equally effective, while direct punishment proves superior under high-cost, high-fine conditions (see region II in Fig.~\ref{Fig_diff}). These findings remain robust across different dilemma strengths, providing valuable insights into the comparative effectiveness of punishment strategies and highlighting the critical role of the cost-to-fine ratio in determining their success within structured populations.

By exploring the comparative effectiveness of direct punishment and indirect punishment, we offer new insights into how different punishment forms promote cooperation in structured populations. Most existing evolutionary models focus on direct punishment because it is unlikely to be evolutionarily stable and requires explanation. Behavioral experiments further support this, showing that individuals who achieve the highest total payoff tend not to use costly (direct) punishment, suggesting this behavior may have evolved for other reasons~\cite{dreber2008winners}. In contrast, a form of indirect punishment---characterized by withholding cooperation---does not face the same issues as direct punishment, such as second-order free riding. Consequently, evolutionary game literature often questions why people continue to use direct punishment when they could opt to withhold cooperation~\cite{ohtsuki2009indirect,panchanathan2004indirect}. 

Moving beyond these evolutionary considerations, we identified a narrow parameter region in structured populations where indirect punishment is more effective than direct punishment. This finding is somewhat surprising because direct punishment directly undermines defectors’ payoffs at the boundaries of cooperative clusters—a critical mechanism for network reciprocity. In contrast, indirect punishment targets defective second-order neighbors, encouraging them to cooperate for the benefit of their direct neighbors, yet it does not directly alter the competitive dynamics at the cluster boundaries. These results highlight the nuanced roles that direct and indirect punishment play in fostering cooperation and suggest that their relative effectiveness is closely linked to the underlying network structure.

This unexpected effectiveness of indirect punishment behaviors, particularly under conditions of low costs and fines, prompts the need for a deep explanation of the mechanisms driving this outcome. We propose a novel explanation: under these conditions, indirect punishment lead to significant accumulated fines on defectors without significantly increasing the punishers' total costs. This is due to the broader punishment scope on indirect neighbors, allowing defectors to be penalized from multiple sources, which amplifies the overall punitive effect, as illustrated in Fig.~\ref{Fig_snap} and Fig.~\ref{Fig_dis1}. In contrast, direct punishment targets defectors within a limited range, reducing its impact. This explanation deepens our understanding of how network reciprocity supports altruistic behavior, traditionally focused on the formation and maintenance of cooperative clusters~\cite{nowak1993spatial,wang2013insight,perc2017statistical}.

In evolutionary game theory, direct punishment refers to a mechanism where an individual actively and intentionally penalizes another for their behavior, typically at a personal cost, such as reducing an immediate neighbor’s payoff in a spatial lattice. Conversely, indirect punishment involves enforcing cooperation through social norms, reputation, or third-party systems, extending influence beyond direct individual action. In our adaptation for a square lattice, we redefine indirect punishment to target second-degree neighbors, capturing its broader spatial scope. While some might note that this retains elements of direct punishment---like individual costs or retaliation risks---these features stem from the lattice’s localized structure and do not undermine the distinction. The critical difference lies in the punishment range: immediate for direct, extended for indirect. This approach preserves the spirit of indirect punishment while enabling a precise analysis of how punishment range influences cooperation in structured populations, enhancing the study of spatial evolutionary dynamics.

Although our adapted form of indirect punishment does not fully align with previous definition—which emphasizes withholding cooperation rather than imposing sanctions\cite{dreber2008winners,rockenbach2006efficient}—our model remains valid as it captures the core concept of sanctioning defectors who harm others. In spatial structures, interactions are limited to local neighbors, prompting us to adapt a novel form of indirect punishment to assess its influence on cooperation in networked populations. Implementing the traditional definition would require higher-order networks to capture group interactions beyond the capabilities of a square lattice~\cite{guo2021evolutionary,civilini2024explosive}, as indirect punishment derived from indirect reciprocity involves ternary or higher-order relationships. Future work could explore higher-order networks to more accurately reflect the dynamics of withholding cooperation. 

Additionally, extending the current model by considering broader punishment scopes beyond indirect neighbors could test the robustness of indirect punishment' effectiveness compared to direct punishment. Such extensions are meaningful because they can quantify the potential future returns for costly punishers on the effectiveness of cooperation and may reveal additional evolutionary mechanisms behind the evolution of costly punishment. Incorporating reinforcement learning algorithms~\cite{hu2024best} or large language models~\cite{ren2024emergence} could also provide a framework for agents to adapt their punishment strategies based on learned experiences, offering insights into how these strategies evolve over time. Finally, considering the competition between different forms of punishment is also meaningful, as the competitive outcomes may help understand various form of punishment can evolve and mechanisms that drive preferences for punishment forms, warrant further investigation.

\paragraph*{Acknowledgments}

We acknowledge support from (i) China Scholarship Council (Grant no.~202308530309) to Z.H., (ii) JSPS KAKENHI (Grant no. JP 23H03499) to C.\,S., and (iii) the grant-in-Aid for Scientific Research from JSPS, Japan, KAKENHI (Grant No. JP 20H02314 and JP 23H03499) awarded to J.\,T.

\paragraph*{Author contributions} 
C.S. and J.T. conceptualised, designed the study, methodology; Y.W, Z.H. and C.S. formal analysis, visualization, validation; Y.W, Z.H., C.S. and J.T. wrote original draft; C.S. and J.T. provided overall project supervision, review and editing.

\paragraph*{Conflict of interest} Authors declare that no conflict of interests.

\paragraph*{Data accessibility} The code used in the study to produce all results is
freely available at: \url{https://github.com/Yujia-WEN/Indirect-reward-outperform-direct-punishment}.

\appendix
\section{}
\label{sec:sample:appendix}
\setcounter{figure}{0}
\renewcommand{\thefigure}{A\arabic{figure}}

\begin{figure}[!ht]
    \centering
    \includegraphics[width=0.95\linewidth]{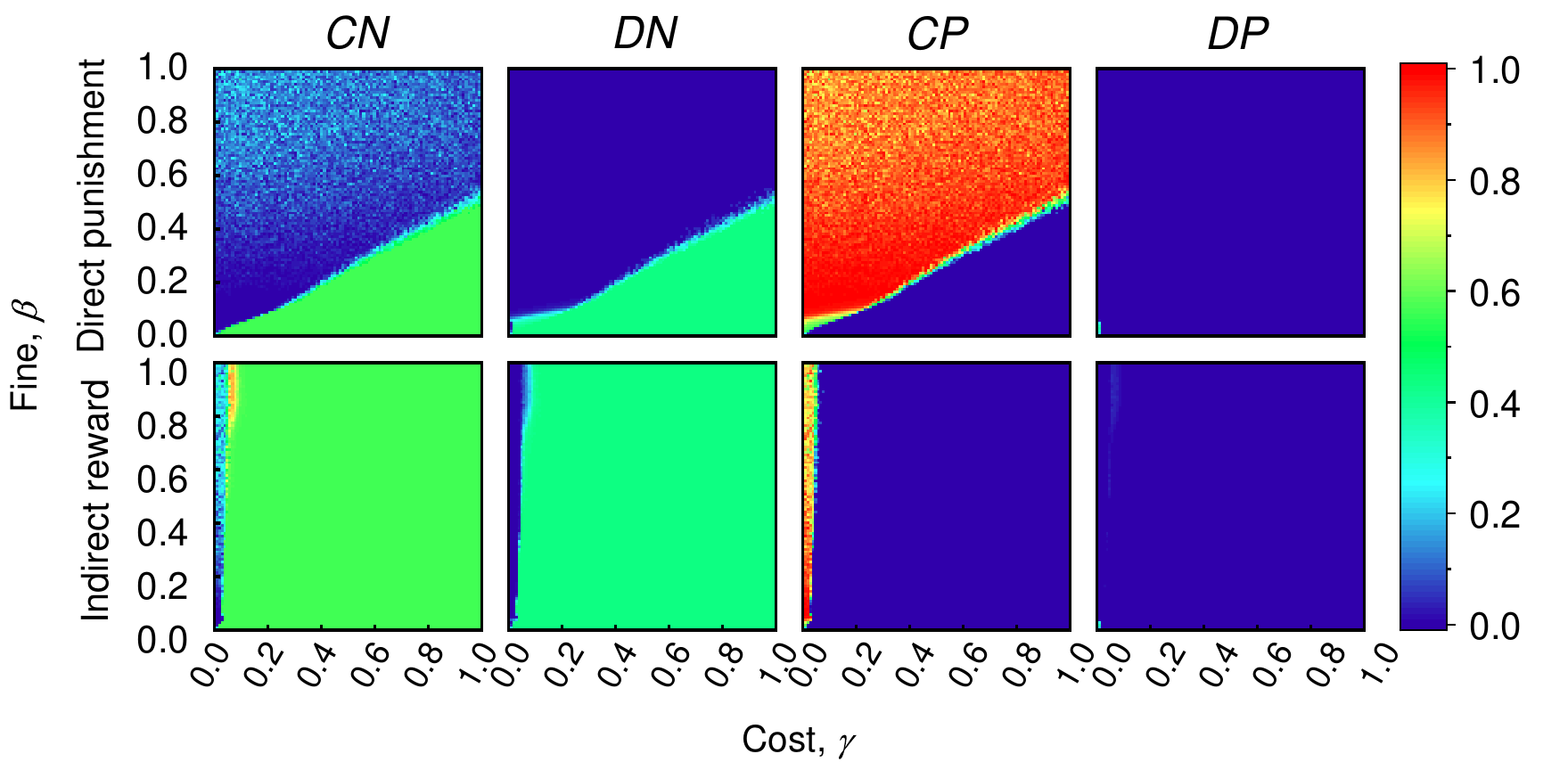}
      \caption{
      Fraction of $CN$, $DN$, $CP$ and $DP$ strategy as functions of punishment cost $\gamma$ and fine $\beta$ under low dilemma strength $r=0.05$ for direct punishment (top panels) and indirect reward (bottom panels).}
      \label{figA1}
\end{figure}

\begin{figure}[!ht]
    \centering
    \includegraphics[width=0.95\linewidth]{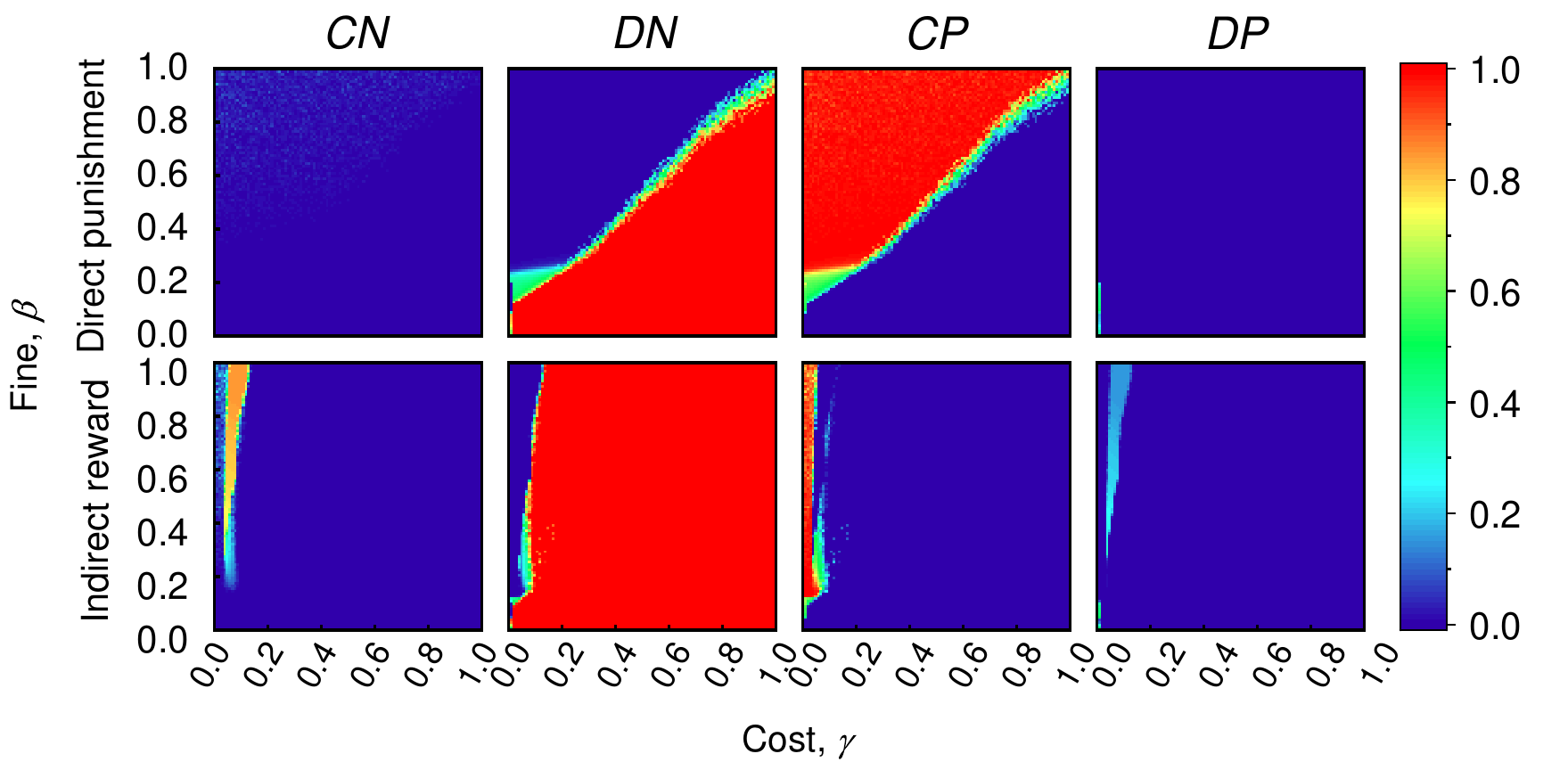}
      \caption{Fraction of $CN$, $DN$, $CP$ and $DP$ strategy as functions of punishment cost $\gamma$ and fine $\beta$ under high dilemma strength $r=0.2$ for direct punishment (top panels) and indirect reward (bottom panels).}
      \label{figA2}
\end{figure}

Fig.~\ref{figA1} demonstrates that, at low dilemma strength, direct punishment can sustain cooperation (i.e., $CN$ and $CP$) across a broad range of $\gamma\text{-}\beta$ parameters. For a fixed punishment cost, an increase in $\beta$ enhances negative feedback from punishments, thus facilitating cooperation. However, for $\beta$, increasing costs impose a burden on the punisher, leading to the disappearance of cooperation In contrast, indirect rewards can only maintain cooperation within a narrow range of $\gamma$, even a slight increase in $\gamma$ can disrupt cooperation. Under high dilemma strength, similar results can be found in Fig.~\ref{figA2}.

\bibliographystyle{elsarticle-num}
\bibliography{biblio}

\end{document}